\documentclass[final,5p,times,twocolumn]{elsarticle}

\usepackage{amssymb}
\usepackage{amsmath}
\biboptions{numbers,sort&compress}
\usepackage{xcolor}

\journal{Journal of Applied Nonlinear Dynamics}

\begin{document}

\begin{frontmatter}



\title{Convergence Dynamics and Scaling Laws in the Dissipative Relativistic Kicked Rotator}
\author{Daniel Borin$^{\rm 1,2}$}
\ead{d.borin@hotmail.com}
\author{Danilo S. Rando$^{\rm 2,1}$}	
\author{Edson D. Leonel$^{\rm 2}$}
\ead{edson-denis.leonel@unesp.br}
\author{Diego F. M. Oliveira$^{\rm 1}$}
\ead{diegofregolente@gmail.com}

	\address{$^{\rm 1}$School of Electrical Engineering and Computer Science, University of North Dakota, Grand Forks, North Dakota 58202, USA. \\
		$^{\rm 2}$São Paulo State University (UNESP), Institute of Geosciences and Exact Sciences, Rio Claro, SP, Brazil}

\begin{abstract}
We investigate the convergence dynamics of this system near period-doubling bifurcations by combining analytical derivations and large-scale numerical simulations. At the bifurcation threshold ($K = K_c$), the dynamics reduce to a normal form that produces a power-law decay $d(n) \propto n^{-1/2}$, from which the critical exponents $\alpha = 1$, $\beta = -1/2$, and $z = -2$ are derived. These analytical predictions are confirmed numerically and shown to satisfy the homogeneous scaling relation $z = \alpha / \beta$. Linearization of the map near the fixed point yields an exponential relaxation law $d_n = d_0 e^{-n/\tau}$ for $K < K_c$, with $\tau \propto (K_c - K)^{-1}$, leading to the relaxation exponent $\delta = -1$. The remarkable agreement between theory and simulation demonstrates that the dissipative relativistic kicked rotator shares the same universality class as one-dimensional unimodal maps, despite its higher dimensionality and relativistic corrections. 
\end{abstract}


\end{frontmatter}




\section{Introduction}

Bifurcations are fundamental to the study of nonlinear dynamical systems \cite{alligood1998chaos,hirsch2013differential}, providing a mathematical framework to understand how qualitative changes in system behavior arise as control parameters are varied \cite{simpson2009simultaneous,strogatz2018nonlinear,oliveira2008feigenbaum}. These transitions mark the boundary between distinct dynamical regimes, such as steady states, periodic oscillations, and chaotic motion, and play a central role in explaining how complex behaviors emerge from simple deterministic rules \cite{hogan2002nonlinear,orel2019non,li2022adaptive}. A single bifurcation event can drastically alter the phase-space structure of a system, creating or annihilating attractors, and changing the number or stability of equilibrium points \cite{oliveira2011dynamical,hansen2014statistical}. 

From a theoretical point of view, bifurcations represent singularities in the parameter space of dynamical equations, where infinitesimal parameter variations produce qualitative shifts in the long-term behavior of trajectories \cite{guckenheimer2013nonlinear,kuznetsov1998elements,oliveira2013some,fm2024mapping}. The classical classification of bifurcations (saddle-node, pitchfork, Hopf, and period-doubling) has been central to the development of modern nonlinear dynamics. Among these, the period-doubling bifurcation stands out as a universal route to chaos, occurring in diverse physical and biological systems and characterized by well-defined scaling properties \cite{strogatz2018nonlinear,govaerts2000numerical,oliveira2008feigenbaum}. Near the bifurcation threshold, the system exhibits self-similar behavior, where the evolution of trajectories can be described by scaling laws and critical exponents, concepts that parallel those found in phase transitions and critical phenomena \cite{BORIN2024114597, borin2024buzz,chen2020scaling,feigenbaum1978quantitative}.

The study of bifurcations extends far beyond abstract mathematics, offering explanatory power across a wide range of scientific disciplines. In physics, bifurcation analysis reveals the mechanisms underlying transitions to turbulence in fluid dynamics \cite{holmes2012turbulence}, the onset of instabilities in plasma systems \cite{escande2016contributions}, and resonance structures in celestial mechanics \cite{valli2013nonlinear}. In biology, it provides information on rhythm generation in neural and cardiac systems \cite{ashwin2024network}, predator–prey oscillations in ecology \cite{caravaggio2018nonlinear}, and regulatory switches that control gene expression \cite{rosenfeld2009characteristics}. In engineering, bifurcation theory enables the design of feedback and control systems capable of stabilizing or harnessing nonlinear dynamics \cite{chen2012bifurcation}. These examples show the interdisciplinary nature of bifurcation theory across the natural and applied sciences.

Within this broader context, the kicked rotator model \cite{levi2003quantum,berry1984incommensurability} serves as a paradigmatic system to explore the emergence of chaos in periodically driven dynamical systems. The model, which represents a rotor subject to periodic impulses, captures the essential features of nonlinear resonance and energy transfer in systems ranging from molecular dynamics to quantum chaos \cite{kaplan1998linear,balazs1986chaos}. When dissipation is introduced, the kicked rotator transitions from an area-preserving to a contracting phase space, leading to the formation of attractors and allowing the study of how dissipative mechanisms influence the route to chaos \cite{bogomolny1988smoothed,oliveira2014statistical}.

In this work, we investigate the dissipative relativistic kicked rotator, a generalized version of the standard map that incorporates relativistic corrections and a tunable dissipation parameter. 
By systematically varying the control parameter $K$, we identify the sequence of period-doubling bifurcations that mark the transition from periodic motion to chaos. At the critical bifurcation point, we examine the convergence dynamics toward stationary states and demonstrate that they obey well-defined scaling laws characterized by universal critical exponents.

To describe these convergence dynamics, we employ a homogeneous and generalized function framework \cite{teixeira2015convergence,oliveira2018scaling}. This approach allows the scaling properties of the stationary-state relaxation to be expressed in terms of universal exponents that remain invariant across distinct dynamical systems. We show that the evolution of the distance from the stationary state follows two distinct regimes, namely, a plateau and a power-law decay, linked through a crossover that satisfies a well-defined scaling relation. Near the bifurcation point, the convergence transitions to an exponential relaxation characterized by a relaxation time $\tau$ that scales to $\tau \propto \mu^{\delta}$, where $\mu$ measures the distance from the critical control parameter.

Our results reveal that the critical exponents obtained numerically, $\beta \approx -1/2$ and $\delta \approx -1$, agree with those predicted for one-dimensional maps and with results observed in other dissipative systems such as the Fermi–Ulam model \cite{s2023bifurcations,leonel2007scaling,oliveira2009scaling}. This agreement highlights the universality of scaling behavior in convergence dynamics, demonstrating that dissipative two-dimensional systems can display the same self-similar properties observed in simpler mappings. 

In addition to the numerical characterization, we present a detailed analytical derivation of the scaling laws derived from the local expansion of the map near the bifurcation point. This analysis clarifies the theoretical origin of the critical exponents and shows that both the power-law and exponential relaxation regimes arise naturally from the normal form of the dissipative relativistic kicked rotator.

The paper is organized as follows. Section~II introduces the model and the governing equations. Section~III investigates the convergence process near a period-doubling bifurcation, focusing on the emergence of scaling laws and the identification of critical exponents. Section~IV presents the analytical validation of the scaling behavior, and Section~V concludes with final remarks.

\section{The model and the mapping}
\label{sec2}

The kicked rotator is one of the most celebrated models in nonlinear dynamics \cite{berry1984incommensurability,levi2003quantum}. Despite its apparent simplicity, this periodically forced system captures the transition from regular motion to chaos and has become a cornerstone in the study of complex dynamics, classical–quantum correspondence, and statistical relaxation \cite{chirikov1979universal,casati2005stochastic,izrailev1990simple,reichel1992transition,ott2002chaos}. Its appeal stems from the fact that a discrete-time mapping derived from simple physical principles reproduces a remarkably rich set of behaviors observed in real-world systems. Examples range from the motion of charged particles in electromagnetic traps to light propagation in nonlinear optical lattices and the diffusion of cold atoms in optical potentials \cite{kaplan1998linear,balazs1986chaos}.

The dynamics of the dissipative relativistic kicked rotator can be written as a two-dimensional mapping acting on the canonical variables $(\theta, I)$, which represent the angular position and the corresponding conjugate momentum at discrete time steps $n \to n+1$. Following the procedure in \cite{oliveira2014statistical}, the map describing the model can be written as:
\begin{equation}
S:\left\{
\begin{array}{ll}
	\theta_{n+1} = [\,\theta_n + \dfrac{I_n}{\sqrt{1+(\rho I_n)^2}} - \xi I_n\,] \ \mathrm{mod}\ 2\pi, \\[0.4cm]
	I_{n+1} = (1 - \psi) I_n + K \sin(\theta_{n+1}),
\end{array}
\right.
\label{eq1}
\end{equation}
where $K$ denotes the strength of the kick, $\rho$ controls the degree of relativistic correction, $\xi$ introduces a velocity-dependent coupling term, and $\psi \in [0,1]$ quantifies the fractional dissipation per kick.

The parameter $K$ drives the transition from integrable ($K=0$) to chaotic dynamics ($K\neq0$) and acts as the main control parameter for bifurcations. The relativistic parameter $\rho$ modulates the nonlinear relation between momentum and velocity: in the Newtonian limit ($\rho\!\rightarrow\!0$), the map reduces to the standard map, while in the ultra-relativistic limit ($\rho\!\rightarrow\!\infty$) the system tends toward integrability again when dissipation is absent \cite{perre2020dynamics}. Dissipation, represented by $\psi$, breaks the area-preserving nature of the system and leads to phase-space contraction, allowing the trajectories to converge towards the attractors. When both $\psi$ and $\xi$ vanish, Eq.~(\ref{eq1}) reduces to the conservative relativistic standard map, for which the phase-space volume is preserved. The complete structure of the parameter space, including the intricate network of shrimp-shaped stability islands, has been extensively explored in \cite{oliveira2014statistical}.

\begin{figure}[!ht]
	\includegraphics[width=1\linewidth]{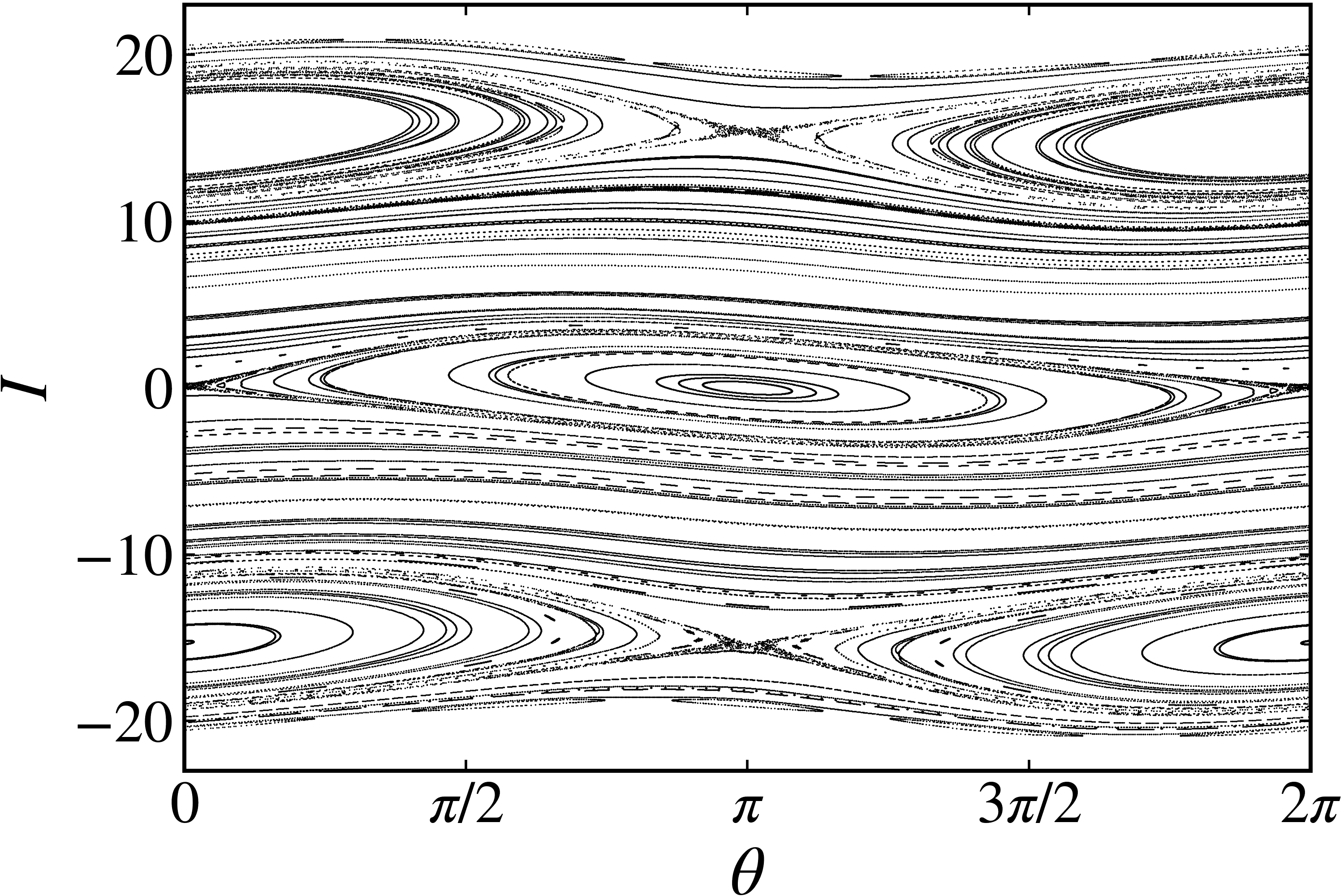}
	\caption{Phase space for $\psi = 0$, $\xi = 0.4$, $\rho = 0.15$, and $K = 1.55$. The conservative dynamics exhibits a mixed structure composed of KAM islands, chaotic regions, and invariant spanning curves.}
	\label{Fig1}
\end{figure}

\medskip
\noindent\textbf{Conservative regime:}  
When $\psi=0$, the system is conservative and the total volume of the phase-space is preserved at each iteration. The resulting dynamics display the well-known coexistence of regular and chaotic domains. Invariant spanning curves, chaotic seas, and families of Kolmogorov–Arnold–Moser (KAM) islands coexist, as shown in Fig.~\ref{Fig1}. Within the islands, trajectories are quasi-periodic and confined to closed orbits, whereas in chaotic regions nearby trajectories diverge exponentially, yet remain bounded by invariant manifolds. This coexistence of regularity and chaos underpins many of the subtle features of Hamiltonian transport, including stickiness and anomalous diffusion \cite{zaslavsky2007physics,zaslavsky2002chaos}.

The conservative kicked rotator serves as an example for the study of universality in chaos. It shares scaling properties with other conservative systems, such as the Fermi–Ulam model and the standard map, enabling the exploration of critical thresholds for the onset of global chaos. However, purely conservative dynamics cannot exhibit asymptotic convergence or stationary attractors, motivating the introduction of controlled dissipation.

\begin{figure}[!ht]
    \centering
    \includegraphics[width=1\linewidth]{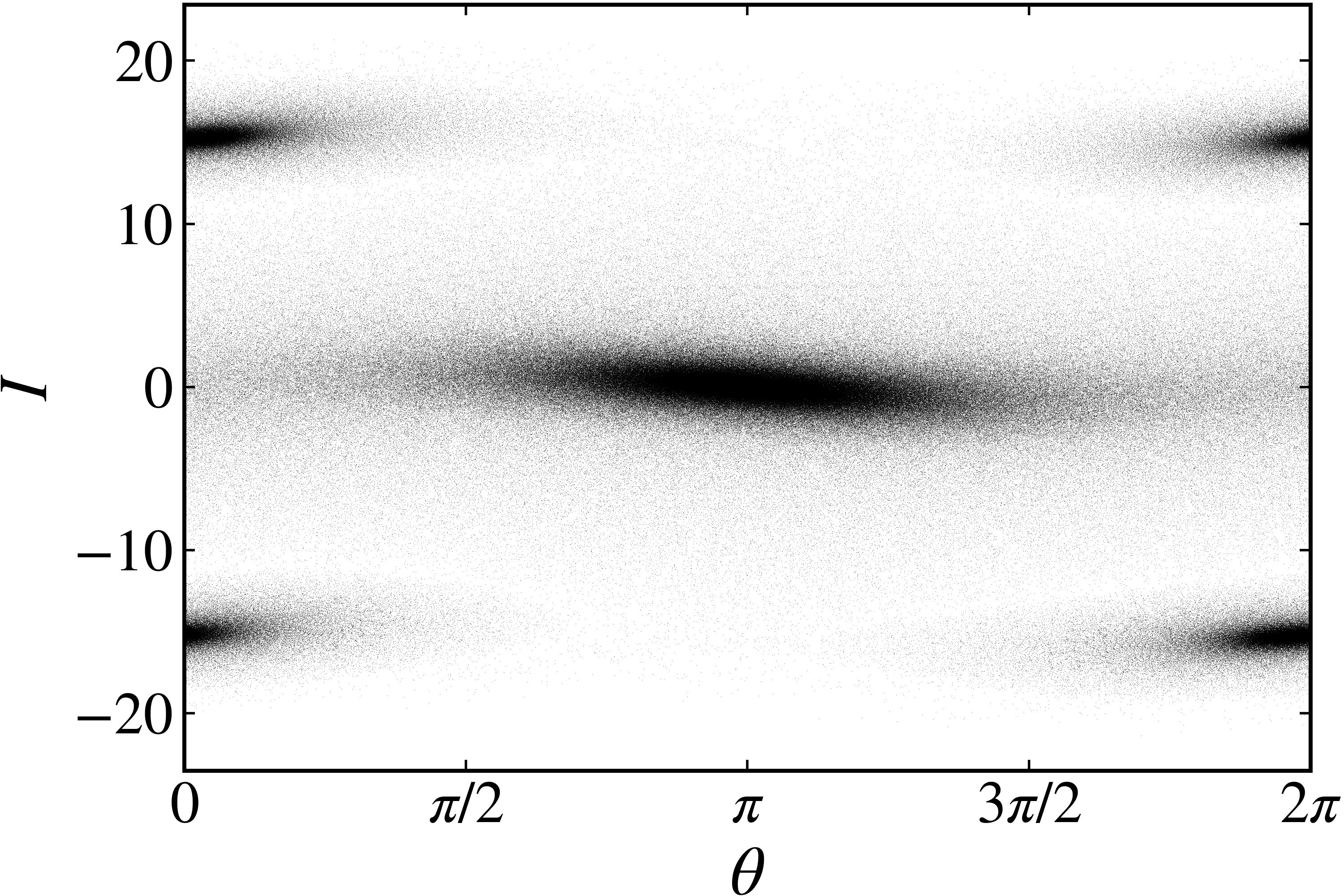}
    \caption{Phase space for the dissipative dynamics with $\psi = 0.01$, $\xi = 0.4$, $\rho = 0.15$, and $K = 1.55$. Dissipation induces phase-space contraction and the emergence of stable periodic attractors.}
    \label{Fig2}
\end{figure}

\medskip
\noindent\textbf{Dissipative regime:}  
When $\psi>0$, the volume of the phase-space contracts exponentially, and trajectories are drawn toward a finite number of periodic or chaotic attractors. By introducing dissipation, the unbounded diffusion of the momentum variable $I$ is halted, and the invariant structures typical of the conservative regime are fundamentally altered. As shown in Fig.~\ref{Fig2}, continuous sets of invariant tori are replaced by isolated periodic attractors that dictate the system’s long-term behavior. This transition from conservative to dissipative dynamics is fundamental for describing real physical systems, where energy dissipation cannot be neglected.

Dissipation introduces new timescales into the dynamics, governing the rate at which trajectories converge to the attractor and altering the spectral properties of the map. The coexistence of multiple attractors, each with its own basin of attraction, can generate fractal basin boundaries, leading to sensitivity to initial conditions even in the absence of chaos \cite{grebogi1989chaos,grebogi1983fractal,aguirre2009fractal}. As the control parameter $K$ increases, these attractors undergo successive bifurcations, most prominently a period-doubling cascade that leads to chaos through a universal route similar to that observed in the logistic and Hénon maps \cite{henon1976two,ott2002chaos} and other systems \cite{oliveira2008feigenbaum,oliveira2013some,fm2024mapping}.

\begin{figure}[!ht]
    \includegraphics[width=1\linewidth]{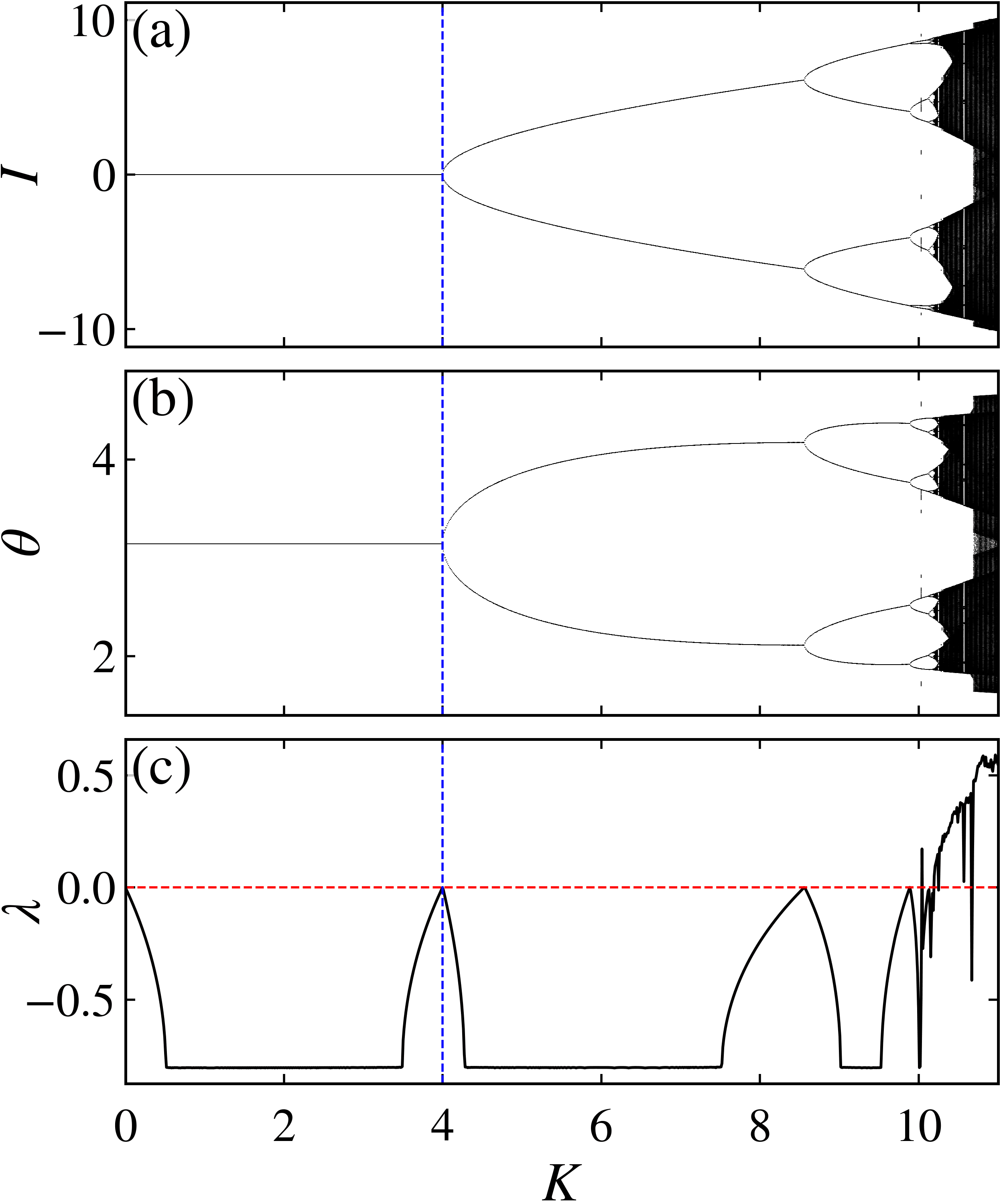}
	\caption{Bifurcation cascade for (a) $I$ and (b) $\theta$ and (c) the lyapunov exponent $\lambda$ as functions of the control parameter $K$, showing the transition from periodic to chaotic dynamics. The blue dashed line indicates the critical parameter $K_c$, where the Lyapunov exponent becomes zero. Parameters: $\psi = 0.8$, $\xi = 0.4$, and $\rho = 0.15$.}
	\label{Fig3}
\end{figure}

To quantify the stability of trajectories and to identify the transition to chaos, we compute the Lyapunov exponents, which measure the exponential rates of divergence or convergence of nearby trajectories in phase space. Figures~\ref{Fig3}(a,b) show the bifurcation diagrams for the variables $I$ and $\theta$ as functions of the control parameter $K$. These diagrams reveal a pitchfork bifurcation, followed by a characteristic sequence of period-doubling bifurcations that eventually lead to chaotic motion. To generate this figure, we consider $10$ random initial conditions within the interval $(\theta_0, I_0) \in [0, 2 \pi] \times [-20, 20]$, evolve them for $10,000$ iterations, discard the first $9,000$ iterations to eliminate transients, and plot the remaining trajectories for each $K \in [0, 11]$. Figure~\ref{Fig3}(c) shows the corresponding Lyapunov exponent for a fixed initial condition $(\theta_0, I_0) = (3, 0.1)$.

The Lyapunov exponents \cite{eckmann1985ergodic} are defined as
\begin{equation}
\lambda_j = \lim_{n \to \infty} \frac{1}{n} \ln |\Lambda_j|, \quad j = 1,2,
\label{eq4}
\end{equation}
where $\Lambda_j$ are the eigenvalues of the matrix $M = \prod_{i=1}^{n} J_i(\theta_i,I_i)$, and $J_i$ denotes the Jacobian of the mapping evaluated along the orbit. A straightforward evaluation of Eq.~(\ref{eq4}) is numerically unstable for large $n$ because the matrix elements of $M$ grow exponentially. To overcome this difficulty, the Jacobian is decomposed at each iteration as $J_i = \Theta_i T_i$, where $\Theta_i$ is an orthogonal matrix and $T_i$ is an upper-triangular matrix obtained through QR decomposition. The Lyapunov exponents are then computed as
\begin{equation}
\lambda_j = \lim_{n \to \infty} \frac{1}{n} \sum_{i=1}^{n} \ln |T_{jj}^{(i)}|, \quad j = 1,2.
\label{eq005}
\end{equation}

These exponents satisfy $\lambda_1 > \lambda_2$, and use the largest Lyapunov exponent, $\lambda_1$, to characterize the system's dynamics.
A positive Lyapunov exponent $\lambda_1$ indicates sensitivity to initial conditions and, consequently, chaotic dynamics, whereas negative values signify convergence to stable attractors. The bifurcation point is characterized by $\lambda_1 = 0$, corresponding to the loss of stability of a periodic orbit and the emergence of a new one with twice the period. An accurate determination of this critical value, $K_c$, is essential for analyzing the scaling and convergence behavior discussed in the following sections. The critical parameter was numerically estimated as $K_c = 3.999999997$  and is shown as the blue dashed line in Figure \ref{Fig3}. This value corresponds to the point at which the largest Lyapunov exponent crosses zero, marking the onset of the bifurcation.

The evolution of the Lyapunov spectrum with respect to $K$ provides a quantitative signature of the route to chaos and a benchmark to verify universality across models. In the dissipative relativistic kicked rotator, as in other nonlinear mappings, the exponent tends smoothly to zero at each period-doubling bifurcation, consistent with the Feigenbaum scenario. The precision of the computed $\lambda_j$ thus not only identifies the bifurcation sequence but also allows an accurate numerical estimation of the critical parameter $K_c$ used in the scaling analysis that follows.

The mapping in Eq.~(\ref{eq1}) offers a fertile setting for investigating scaling laws near bifurcation points because its dynamics are both nonlinear and dissipative
. As the system approaches the critical value $K_c$, the trajectories exhibit self-similar convergence toward the stationary state. This behavior manifests in the scaling of distance measures, convergence rates, and relaxation times, which are later shown to obey universal power laws characterized by the critical exponents $\alpha$, $\beta$, $z$, and $\delta$. The detailed exploration of these scaling relationships constitutes the central analysis of this work.

\section{Convergence to the stationary state}
\label{sec3}

Understanding how trajectories converge to stationary states provides key insight into the stability and universality of dissipative dynamical systems \cite{hirsch1988stability,lasalle1976stability}. The process of convergence determines not only whether a system reaches equilibrium but also the rate and scaling form by which transient dynamics decay \cite{10.1063/5.0233700}. Near bifurcation points, convergence is nontrivial: small perturbations of the control parameter can dramatically alter relaxation times and generate scaling behavior analogous to critical slowing down in phase transitions \cite{drazin1992nonlinear,feigenbaum1978quantitative}.

For the dissipative relativistic kicked rotator, convergence toward the stationary state occurs in the two-dimensional $(I,\theta)$ phase plane. To quantify this process, we define the Euclidean distance between the instantaneous state $(I_n,\theta_n)$ and the fixed point $(I^{*},\theta^{*})$ as
\begin{equation}
d(n)=\sqrt{(I_n-I^{*})^{2}+(\theta_n-\theta^{*})^{2}}.
\label{eq_d}
\end{equation}
The coordinates $(\theta^{*}, I^{*})$ are determined at the critical parameter $K=K_c$, where the largest Lyapunov exponent vanishes, leading to $(\theta^{*}, I^{*}) = (\pi, 0)$. The function $d(n)$ thus provides a scalar measure of how far the system is from the stationary state at iteration $n$ and serves as an observable for characterizing relaxation dynamics.

\begin{figure}[!ht]
\includegraphics[width=1\linewidth]{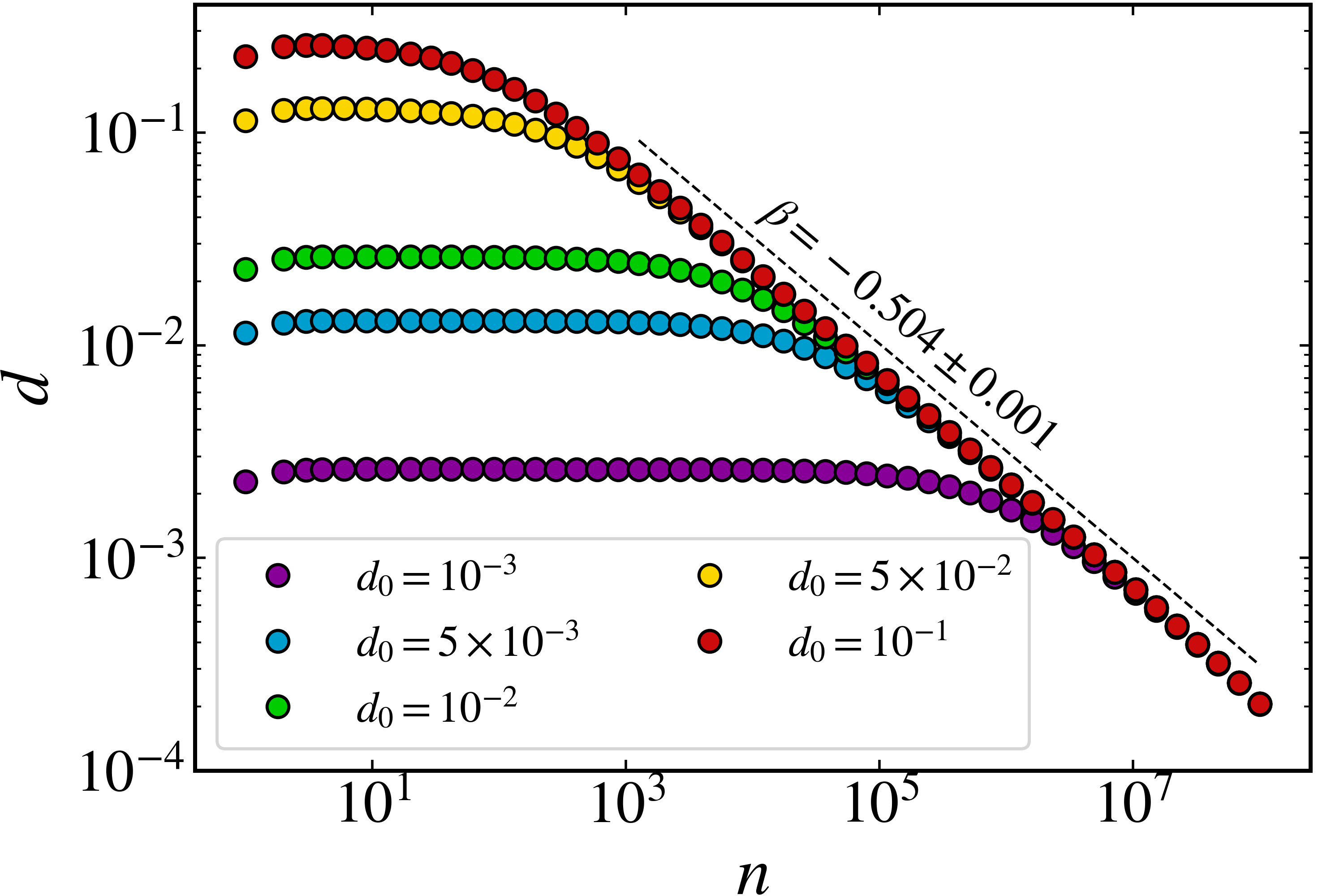}
\caption{Behavior of $d(n)$ for different initial distances $d_{0}$ from the bifurcation point. A plateau regime ($n < n_{x}$) is followed by a power-law decay ($n > n_{x}$) with exponent $\beta \approx -1/2$, confirming the scaling relation $n_{x} \propto d_{0}^{\,z}$.}
\label{Fig4}
\end{figure}

Figure~\ref{Fig4} displays the behavior of $d(n)$ for a set of initial distances $d_{0}$ spanning several orders of magnitude. At early times, the curves exhibit an initial plateau, followed by a transition to a faster decay regime, characterized by a power-law with slope $\beta = -0.504 \pm 0.001$.

To describe convergence, we employ a homogeneous and generalized function formalism \cite{teixeira2015convergence,oliveira2018scaling}. The central hypothesis is that, near the bifurcation, the observable $d(n,d_{0})$ obeys a scaling relation of the form
\begin{equation}
d(n,d_{0}) = \ell\, d(\ell^{a} n,\ell^{b} d_{0}),
\label{homog}
\end{equation}
where $\ell$ is a scaling factor, and $a$ and $b$ are characteristic exponents describing how time $n$ and the initial deviation $d_{0}$ scale relative to each other. This equation expresses the self-similarity of the convergence dynamics: rescaling time and initial conditions by powers of $\ell$ leaves the functional form invariant up to a multiplicative factor.

Two scaling limits can be extracted from Eq.~(\ref{homog}).  
For short times ($n \ll n_{x}$), the dynamics is still governed by local behavior near the initial condition, and $d(n)$ remains approximately constant:
\begin{equation}
d(n)\propto d_{0}^{\alpha}, \qquad n \ll n_{x},
\label{eq_alpha}
\end{equation}
where $\alpha$ is the plateau exponent.  
At longer times ($n \gg n_{x}$), the transient decay dominates, and the system follows a power-law approach to the stationary state:
\begin{equation}
d(n)\propto n^{\beta}, \qquad n \gg n_{x},
\label{eq_beta}
\end{equation}
where $\beta$ characterizes the asymptotic decay. The crossover iteration $n_{x}$, shown in Figure~\ref{Fig5}(a), between these regimes scales with the initial distance as
\begin{equation}
n_{x} \propto d_{0}^{\,z}, \qquad z = \frac{\alpha}{\beta}.
\label{eq_z}
\end{equation}
Equations~(\ref{eq_alpha})–(\ref{eq_z}) collectively define a scaling law that encapsulates the universal features of convergence.  Similar relations have been reported for one-dimensional dissipative maps and for the Fermi–Ulam model \cite{oliveira2009scaling,leonel2007scaling}, demonstrating the cross-system robustness of these scaling exponents.

\begin{figure*}[!ht]
\includegraphics[width=1\linewidth]{Figures/fig5_new_lite.png}
\caption{
(a) Method for determining the crossover time $n_x$. The plot shows $d(n)$ along with two additional curves: one representing the initial plateau $d_{ip}$ and the other describing the power-law decay. The crossover time $n_x$ is defined as the point where these two curves intersect.
(b) Determination of the plateau exponent $\alpha$. The initial plateau $d_{ip}$ is plotted as a function of the initial distance $d_0$, and a power-law fit is performed, resulting in $\alpha = 0.996 \pm 0.002$.
(c) Determination of the crossover exponent $z$. The values of $n_x$ are plotted against $d_0$, and a power-law fitting yields the exponent $z = -1.978 \pm 0.003$.
(d) Scaling collapse of the convergence curves. After applying the transformations 
$n \rightarrow n / d_0^{z}$ and $d \rightarrow d / d_0^{\alpha}$, the curves describing the convergence dynamics collapse onto a single universal curve, confirming the scaling invariance of the observable.
}
\label{Fig5}
\end{figure*}

Initially, we determine these exponents through numerical analysis. The plateau exponent $\alpha$ is obtained from the plot of the initial plateau $d_{ip}$ as a function of $d_0$, following Equation~\eqref{eq_alpha}, using a power-law fitting. This analysis is shown in the Figure~\ref{Fig5}(b), leading to $\alpha = 0.996 \pm 0.002$. The exponent $\beta$ was determined in Figure~\ref{Fig4}, by  fitting the decay regime, giving $\beta = -0.504 \pm 0.001$. The crossover exponent $z$, which is characterized by Equation~\ref{eq_z}, is obtained from the plot of $n_x$ versus $d_0$, shown in Figure~\ref{Fig5}(c). To determine $n_x$, we plot two curves: one describing the initial plateau $d_{ip}$ and another representing the power law decay. The crossover interaction $n_x$ corresponds to the intersection of these two curves, as illustrated in Figure~\ref{Fig5}(a). From the fit shown in Figure~\ref{Fig5}(c), we obtain the crossover exponent $z=-1.978 \pm 0.003$. Then, under the transformations $n \rightarrow n / d_0^z$ and $d \rightarrow d / d_0^\alpha$, the curves describing the convergence overlap onto a single universal curve, as shown in Figure~\ref{Fig5}(d), confirming the scaling invariance of the observable.


The observed exponents have clear physical interpretations. The plateau exponent $\alpha$ captures the short-time self-similarity displayed by trajectories before nonlinear effects become significant, whereas $\beta$ measures how quickly deviations decay once the nonlinear terms take over. The crossover exponent $z$ determines how strongly the duration of the transient depends on the initial condition, analogous to the divergence of correlation lengths near critical points in statistical physics. The close agreement between the numerically measured exponents and the theoretical prediction $z = \alpha/\beta$ highlights the internal consistency of the scaling description.

After analyzing the system at $K=K_c$, we now turn our attention to studying the convergence for $K$ values slightly smaller than $K_c$.
As the system approaches the critical control parameter $K_{c}$, the power-law regime gives way to an exponential decay. In this near-critical regime, the convergence to the stationary state can be described by
\begin{equation}
d(n)=d_{0}\, e^{-n/\tau},
\label{d_expo}
\end{equation}
where $\tau$ denotes the relaxation time.  The relaxation time diverges as the system approaches the bifurcation according to
\begin{equation}
\tau \propto \mu^{\,\delta}, \qquad \mu = K_{c}-K,
\label{eq_tau}
\end{equation}
with $\delta$ the relaxation exponent. This expression captures the phenomenon of critical slowing: as $K\to K_{c}$, trajectories require increasingly longer times to settle into the stationary state.

\begin{figure}[!ht]
\centering
\includegraphics[width=1\linewidth]{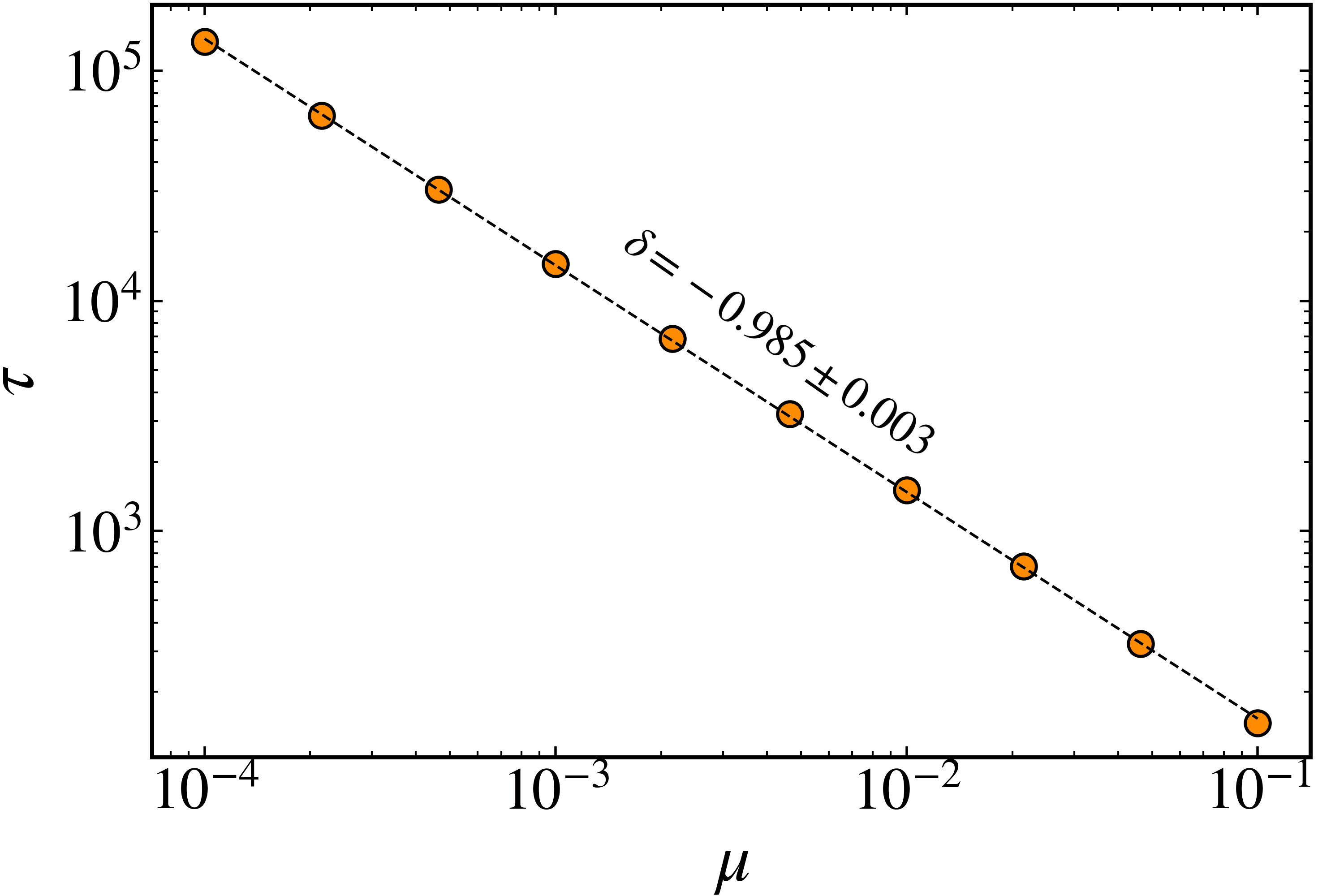}
\caption{Relaxation time $\tau$ as a function of $\mu = K_{c}-K$ for $\psi = 0.8$, $\xi = 0.4$, and $\rho = 0.15$. The power-law fit yields $\delta = -0.985 \pm 0.003$, confirming the universal scaling $\tau \propto \mu^{\delta}$ with $\delta \approx -1$.}
\label{Fig6}
\end{figure}

To determine $\delta$, we perform extensive numerical simulations for $10^{6}$ randomly chosen initial conditions within the basin of attraction (in our case, inside a circle of radius $r=10^{-1}$ centered at fixed point $(\theta^*, I^*)$). For each initial condition, we measure the number of iterations required for $d(n)$ to fall below a tolerance of $10^{-6}$, which we define as the relaxation time. Averaging over all realizations gives the mean relaxation time $\tau$ for each $\mu$. Figure~\ref{Fig6} shows $\tau$ as a function of $\mu$ for fixed parameters $\psi = 0.8$, $\xi = 0.4$, and $\rho = 0.15$. The power-law fit gives $\delta = -0.985 \pm 0.003$.

\section{Analytical validation of the scaling behavior}
\label{sec:analytical_validation}

The numerical results discussed in the previous section reveal a clear scaling structure in the convergence dynamics, characterized by the critical exponents $\alpha = 1$, $\beta = -1/2$, and $z = -2$. To confirm that these relationships emerge from the intrinsic properties of the map rather than from empirical fitting, we now demonstrate their analytical origin based on the local dynamics of the dissipative relativistic kicked rotator near the bifurcation point. The numerical results in Sec.~\ref{sec3} revealed that the distance
\begin{equation}\label{eq:dn}
d(n) = \sqrt{(I_n - I^{*})^2 + (\theta_n - \theta^{*})^2}
\end{equation}
between the orbit and the fixed point exhibits three key features near the first period--doubling bifurcation: (i) an initial plateau whose height scales with the initial distance $d_0$; (ii) a power--law decay $d(n) \propto n^{\beta}$ at the bifurcation threshold; and (iii) an exponential decay with a relaxation time $\tau$ that diverges as $K \to K_c$. In this section we show how these properties follow from the local normal form of the map and how they lead to the exponents
\[
\alpha = 1, \qquad \beta = -\tfrac{1}{2}, \qquad z = -2, \qquad \delta = -1.
\]

Our analysis proceeds in three steps. First, we derive the normal form for the dynamics near the bifurcation point, focusing on the local behavior of the system. Second, we use the normal form to derive the critical exponents $\alpha$, $\beta$ and $z$. Finally, we compute the divergence of the relaxation time $\tau$ for $K < K_c$ and show that $\tau \propto (K_c - K)^{-1}$.

\subsection{Local normal form near the Pitchfork bifurcation}

From the bifurcation diagram exhibited in Figure \ref{Fig3}(a) and (b), we observe that both variables, $\theta$ and $I$, exhibit a pitchfork bifurcation at $K = K_c$. At this point, the fixed point $(\theta^*, I^*)$ loses stability, and two new fixed points emerge. Each of these new fixed points evolves independently, undergoing period-doubling bifurcations until the system reaches chaos.  
Near the critical parameter $K_c$, each variable can be approximated by the normal form of the pitchfork bifurcation:
\begin{equation}
\theta_{n+1} \approx \theta_n + (K-K_c)\,\theta_n - \theta_n^3 + \theta^*
\end{equation}
\begin{equation}
I_{n+1} \approx I_n + (K-K_c)\,I_n - I_n^3 + I^*
\end{equation}
for $K$ near $K_c$.  

Since, aside from the positions of their respective bifurcations, the variables behave similarly, we can approximate them as practically identical near the critical parameter. Defining a single representative variable, we have:
\begin{equation}
\theta_{n+1} - \theta^* \approx I_{n+1} - I^* \approx x_{n+1} = x_n + (K-K_c)x_n - x_n^3
\end{equation}
for $K$ near $K_c$.  
Accordingly, the distance from the fixed point can be expressed as
\begin{equation}
d_{n+1} = \sqrt{(I_{n+1} - I^*)^2 + (\theta_{n+1} - \theta^*)^2} 
\approx \sqrt{2}\,|x_{n+1}| 
\end{equation}
that is,
\begin{equation}
    d_{n+1}, \propto x_{n+1}.
\end{equation}
Thus, we can write $d$ in its normal form:
\begin{equation}\label{eq:normal_d}
d_{n+1} = d_n - \mu d_n - d_n^3,
\end{equation}
where we define $\mu = K_c - K$ and this form is valid for small values of $\mu$ (i.e., when $K$ is close to $K_c$).

\subsection{Analytical determination of the critical exponents $\alpha$, $\beta$, and $z$}

At the bifurcation threshold $K = K_c$, we have $\mu = 0$, and Eq.~\eqref{eq:normal_d} reduces to
\begin{equation}
d_{n+1} = d_n - d_n^3.
\label{eq:J_crit}
\end{equation}
In the continuum limit,
\begin{equation}\label{eq:contlim}
    d_{n+1} - d_n = \frac{d_{n+1} - d_n}{(n+1) - n} \approx \frac{\partial d}{\partial n},
\end{equation}
which leads to the approximation
\begin{equation}
\frac{\partial d}{\partial n} = -d^3.
\label{eq:d_ode}
\end{equation}
The solution of Eq.~\eqref{eq:d_ode} is
\begin{equation}
    d(n) = \frac{d_0}{\sqrt{2 d_0^{2} n + 1}}.
    \label{eq:d_mu0}
\end{equation}

We now analyze the implications of Eq.~\eqref{eq:d_mu0} for different ranges of $n$.  
First, consider the regime $2 d_0^{2} n \ll 1$, equivalent to $n \ll n_x$. In this limit,
\begin{equation}
d(n) \approx d_0.
\end{equation}
A direct comparison with the first scaling hypothesis [Eq.~\eqref{eq_alpha}] shows that the critical exponent is $\alpha = 1$. This result simply reflects that, at early times, the distance to the fixed point remains essentially proportional to the initial displacement.

Next, for longer times, we consider the regime $2 d_0^{2} n \gg 1$, corresponding to $n \gg n_x$. In this limit,
\begin{equation}
d(n) \propto n^{-1/2}.
\end{equation}
Comparison with the second scaling hypothesis [Eq.~\eqref{eq_beta}] then yields the exponent $\beta = -1/2$.

Finally, the crossover iteration $n_x$ between the plateau and power–law regimes is obtained by matching the two asymptotic behaviors, i.e., by setting $2 d_0^{2} n = 1$, or equivalently $n = n_x$. This gives
\begin{equation}
n_x = \frac{1}{2 d_0^{2}} \propto d_0^{-2}.
\end{equation}
Comparison with the third scaling hypothesis [Eq.~\eqref{eq_z}] therefore leads to $z = -2$.  
The relation $z = \alpha / \beta$ follows directly from the homogeneous scaling form \eqref{eq_z}, and the agreement with the numerically measured exponents provides a consistency check for the scaling description.

\subsection{Relaxation time and the exponent $\delta=-1$ for $K<K_c$}

We now consider the case \( K < K_c \), which corresponds to \( \mu > 0 \). In this regime, the normal form of \( d \) is given by Eq.~\eqref{eq:normal_d}:

\begin{equation}
d_{n+1} = d_n - \mu d_n - d_n^3.
\end{equation}

Taking the continuum limit, we obtain the approximation described in Eq.~\eqref{eq:contlim}, which leads to the differential equation

\begin{equation}\label{eq:d_ode_2}
\frac{\partial d}{\partial n} = - \mu d - d^3.
\end{equation}

To investigate the behavior of \( d \) near the fixed point (i.e., the asymptotic state), we assume that \( d \) is very small, so that \( d \to 0 \). In this case, the \( d^3 \) term decays much faster than \( d \), and we can safely neglect the cubic term. With this approximation, the equation simplifies to

\begin{equation}\label{eq:d_ode_3}
\frac{\partial d}{\partial n} = - \mu d.
\end{equation}

The solution to this linear differential equation is

\begin{equation}\label{eq:d_exp_relax}
d(n) = d_0 e^{-\mu n}.
\end{equation}
Comparing this solution to the standard exponential relaxation form
\begin{equation}
d(n) \sim e^{-n/\tau},
\end{equation}
we can identify the relaxation time \( \tau \) as
\begin{equation}
\tau = \mu^{-1}.
\end{equation}
Thus, as the control parameter \( \mu \) approaches the bifurcation point from below, the relaxation time diverges according to a power law:

\begin{equation}
\tau \propto \mu^{\delta}, \quad \delta = -1.
\end{equation}
\label{eq:delta_res}

The exponent \( \delta = -1 \) characterizes the critical slowing down of the convergence rate as the system nears the bifurcation point. This result matches the well-known exponent observed in one-dimensional dissipative maps, indicating that the same underlying mechanism governs the decay of the control parameter near criticality.

Table~\ref{tab_exponents} presents the analytical and numerical values of all critical exponents, showing excellent quantitative agreement. The minor deviations arise from numerical precision limits, finite-size effects, and the discrete sampling of the parameter space, all of which are expected in simulations of this type.

The analytical derivations show that the dissipative relativistic kicked rotator obeys the same normal-form structure as one-dimensional unimodal maps near a period-doubling bifurcation. As a result, the critical exponents obtained here coincide with those reported for the dissipative Fermi--Ulam model~\cite{s2023bifurcations}. This concordance reinforces the universality of the scaling laws, demonstrating that they emerge from the geometric properties of the bifurcation rather than the specific details of the system. The strong agreement between theory and simulation confirms the validity of the scaling framework and highlights the robustness of the homogeneous-function approach for describing convergence dynamics near critical points.

\begin{table}[t]
\centering
\caption{Comparison between analytically predicted and numerically obtained exponents for the dissipative relativistic kicked rotator.}
\label{tab_exponents}
\begin{tabular}{lccc}
\hline
\textbf{Exponent} & \textbf{Analytical prediction} & \textbf{Numerical value}\\
\hline
$\alpha$ & $1$ & $0.996 \pm 0.002$\\
$\beta$ & $-1/2$ & $-0.505 \pm 0.001$\\
$z$ & $-2$ & $-1.978 \pm 0.003$\\
$\delta$ & $-1$ & $-0.985 \pm 0.003$\\
\hline
\end{tabular}
\end{table}

\section{Discussions and conclusions}
\label{sec5}

In this work, we conducted a detailed investigation of the convergence dynamics in the dissipative relativistic kicked rotator, focusing on how trajectories approach stationary states near period-doubling bifurcations. By combining analytical derivations with extensive numerical simulations, we demonstrated that the convergence process follows universal scaling laws characterized by a consistent set of critical exponents. 

Our analysis began with the computation of Lyapunov exponents, which allowed for precise identification of the bifurcation sequence and the critical control parameter $K_{c}$. At this threshold, the evolution of the distance to the stationary state, $d(n)$, was shown to be self-similar and governed by a homogeneous and generalized scaling function. This framework yielded the exponents $\alpha = 0.996 \pm 0.002 \approx 1$, $\beta = -0.505 \pm 0.001 \approx -1/2$, and $z = -1.978 \pm 0.003 \approx -2$, which satisfy the scaling law $z = \alpha / \beta$. In the vicinity of the bifurcation, for $K < K_c$, the relaxation time diverges as $\tau \propto \mu^{\delta}$ with $\delta = -0.985 \pm 0.003 \approx -1$, showing that relaxation becomes progressively slower near the value of the critical parameter.

The analytical results clarify why the numerical scaling laws arise so robustly in the dissipative relativistic kicked rotator. By reducing the dynamics near the bifurcation to the normal form, we showed that the convergence to the fixed point is governed solely by this local normal form. This reduction explains the scaling behavior of \( d(n) \) at \( K = K_c \) and \( K < K_c \), yielding the critical exponents \( \alpha = 1 \), \( \beta = -1/2 \), \( z = -2 \), and \( \delta = -1 \). The homogeneous-function formalism provides a unified framework for this structure: it shows that the scaling relations follow directly from the symmetry of the normal form, making them independent of the global dynamics or relativistic corrections in the full model.

From a theoretical point of view, the emergence of these exponents reflects the fundamental self-similarity of trajectories near the loss of stability. The homogeneous function formalism offers a unified framework that naturally links analytical derivations to numerical scaling laws, demonstrating that the critical behavior depends primarily on the local curvature of the map near the bifurcation, rather than on its global complexity. This invariance explains why systems as diverse as the logistic, Hénon, Fermi–Ulam, and dissipative relativistic kicked rotator models display identical critical exponents. 

\section*{Acknowledgements}
 D.B. state that this study was financed, in part, by the São Paulo Research Foundation (FAPESP), Brasil, Process Numbers \#2024/06749-8 and \#2022/03612-6. D.S.R. gratefully acknowledges the support of CAPES (Brazil) and the University of North Dakota for their hospitality. E.D.L. acknowledges support from the National Council for Scientific and Technological Development (CNPq, Grants 301318/2019-0 and 304398/2023-3) and from the São Paulo Research Foundation (FAPESP, Grants 2021/09519-5 and 2019/14038-6).

\end{document}